\documentstyle[12pt,twoside,psfig]{article}
\begin{document}
\begin{center}{\Large\bf Generalized Second Law of Thermodynamics for Non-canonical Scalar Field Model with Corrected-Entropy}
\\[15mm]
Sudipta Das\footnote{E-mail:  sudipta.das@visva-bharati.ac.in},~ Ujjal Debnath\footnote{E-mail: ujjaldebnath@gmail.com},~
Abdulla Al Mamon\footnote{E-mail : abdullaalmamon.rs@visva-bharati.ac.in}\\
{\em $^{1,3}$Department of Physics, Visva-Bharati,\\
Santiniketan- 731235, ~India.}\\
{\em $^{2}$Department of Mathematics, Indian Institute of Engineering\\
Science and Technology, Shibpur, Howrah-711 103, ~India.}\\
[15mm]
\end{center}
\vspace{0.5cm}
\vspace{0.5cm}
\pagestyle{myheadings}
\newcommand{\be}{\begin{equation}}
\newcommand{\ee}{\end{equation}}
\newcommand{\bea}{\begin{eqnarray}}
\newcommand{\eea}{\end{eqnarray}}
\newcommand{\bc}{\begin{center}}
\newcommand{\ec}{\end{center}}
\begin{abstract}
 In this work, we have considered a non-canonical scalar field dark energy model in the framework of flat FRW background. It has also been assumed that the dark matter sector interacts with the non-canonical dark energy sector through some interaction term. Using the solutions for this interacting non-canonical scalar field dark energy model, we have investigated the validity of generalized second law (GSL) of thermodynamics in various scenarios using first law and area law of thermodynamics. For this purpose, we have assumed two types of horizons viz apparent horizon and event horizon for the universe and using first law of thermodynamics, we have examined the validity of GSL on both apparent and event horizons. Next, we have considered two types of entropy-corrections on apparent and event horizons. Using the modified area law, we have examined the validity of GSL of thermodynamics on apparent and event horizons under some restrictions of model parameters.
\end{abstract} 
\section{Introduction}
Many cosmological experiments, such as observations of type Ia supernovae \cite{Riess,Perl}, large scale structure formation \cite{Teg,Abaz} and cosmic microwave background anisotropy
\cite{Sper1,Sper2} indicate that at present our universe is undergoing a phase of cosmic acceleration. The main source responsible for this accelerated expansion is dubbed as {\em dark energy} (DE) having sufficient negative pressure. A number of candidates have appeared in the literature which serve as a very good candidate for this observed cosmic acceleration. For
example, a cosmological constant (with $w=-1$), dynamical dark energy models (with varying $w$), $f(R)$ gravity models and many other such models turn out to be a good driver of this accelerated expansion. The simplest candidate as the dynamical dark energy can be realized by scalar field models which arise in string theory and are studied as candidate for dark energy. For a review on dark energy models, one can look at \cite{quin, martin, Tsuji}.  However, for most of the dark energy models the analysis is carried out for canonical scalar fields. But there is another class of scalar field models, viz, non-canonical scalar field models which have also been considered as a candidate for dark energy (for details see \cite{sdas2, Fang, unni} and the references there in). This class of models has also been found to be a very
good candidate for dark energy and could provide solution to a number of cosmological problems.\\
\par Usually, most of the dark energy models consider the field to be non-interacting. Presently we live in an epoch where the densities of the dark energy and the dark matter are comparable and this coincidence problem can be easily dealt with by taking a suitable interaction between the two dark sectors. Also the transition from matter domination to dark energy domination can be explained through an appropriate energy exchange rate. Usually, to obtain a suitable evolution of the universe an interaction term is assumed \cite{Berg,Ca,Zim,Hu} where  the decay rate is considered to be proportional to the present value of the Hubble parameter for good fit to the expansion history of the Universe. However, a number of various other forms of interaction have also been considered in which the interaction term has been chosen phenomenologically. Since nothing is known about the nature of DE, search is still on to find a suitable and observationally viable interaction term. So, motivated by these facts, recently non-canonical scalar field DE models are being studied which interacts with the dark matter sector through a source term \cite{msami}. Recently it has been shown that BOSS (Baryon Oscillation Spectroscopic Survey) data also provides evidence for interacting Dark Energy \cite{boss}.\\
\par In the semi-classical quantum description of black hole physics, it was found that black hole emits Hawking radiation with a temperature proportional to their surface gravity at the event horizon with an entropy proportional to its horizon area \cite{Haw,Bek}. The Hawking temperature and horizon entropy together with the black hole mass obey the first law of
thermodynamics \cite{Bar}. In Einstein gravity, Jacobson \cite{Jacob} first derived the Einstein equations from the proportionality of entropy and horizon area together with the
first law of thermodynamics $\delta Q=TdS$ in the Rindler space-time where $\delta Q$ and $T$ are interpreted as the energy flux and temperature seen by an accelerated observer inside the
horizon. For general static spherically symmetric space-time, using Einstein field equation, Padmanabhan \cite{Pad} formulated the first law of thermodynamics on the horizon. The study on the relation between the Einstein equation and the first law of thermodynamics has been generalized to the cosmological context where it was shown that the first law of thermodynamics on the apparent horizon can be derived from the Friedmann equation and vice versa if we take the Hawking temperature $T_{A}=\frac{1}{2\pi R_{A}}$ and the entropy $S_{A}=\frac{A}{4G}$ on the apparent horizon \cite{Cai0} where $A=4\pi R_{A}^{2}$ is the surface area of the apparent horizon and $R_{A}$ is the radius of the apparent horizon.\\
\par By considering the Universe as a thermodynamical system, Gibbons and Hawking \cite{Gibb} first investigated the thermodynamics in de-Sitter space-time. It has been found that in spatially flat de-Sitter space-time, the event horizon and the apparent horizon of the Universe coincide and there is only one cosmological horizon, but in the standard big bang model the cosmological event horizon does not exist. However, for accelerating universe dominated by dark energy, the cosmological event horizon and the apparent horizon are distinct. Besides the first law of thermodynamics, a lot of attention has been paid to the generalized second law (GSL) of thermodynamics in the context of dark energy. Following first law, it has been considered that, at the event horizon, the Hawking temperature is given by $T_{E}=\frac{1}{2\pi R_{E}}$ and the entropy is given by $S_{E}=\frac{A}{4G}$, where $A=4\pi R_{E}^{2}$ is the surface area of the event horizon and $R_{E}$ is the radius of the event horizon. Wang et al \cite{Wang0} has found that the first law and GSL of thermodynamics hold on the apparent horizon, but they break down
for the event horizon. Till now a lot of work \cite{Set0,Iz,Frol,Gong,Shey,Wang1,Ak,Sad} has been done where the validity of first law and GSL of thermodynamics on apparent and event horizons have been investigated in various cosmological phenomena.\\
\par In recent years, it has been shown that the quantum corrections (namely logarithmic and power law corrections) provided to the semi-classical entropy-area relationship leads to the curvature correction in the Einstein-Hilbert action and vice versa \cite{Zhu,Cai00}. The
logarithmic corrections arises from loop quantum gravity due to thermal equilibrium fluctuations and quantum fluctuations \cite{Meis,Jamil1,Sad1}. The general type of Entropy-area relation in logarithmic correction is given by
\begin{equation}\label{log1}
S_{BH}=\frac{A}{4G}+\alpha\ln \frac{A}{4G}+\beta
\frac{4G}{A}+\gamma
\end{equation}
where $\alpha,~\beta$ and $\gamma$ are dimensionless constants of
order unity. On the other hand, power law corrections to the entropy-area relation appear while dealing with the entanglement of quantum fields inside and outside the horizon
\cite{sd3c,She,Wei2} which is given by
\begin{equation}\label{pw1}
S_{BH}=
\frac{A}{4G}\left[1-K_{\lambda}A^{1-\frac{\lambda}{2}}\right],
\end{equation}
where, $
K_{\lambda}=\frac{\lambda(4\pi)^{\frac{\lambda}{2}-1}}{(4-\lambda)r_{c}^{2-\lambda}}
$ and $r_{c}$ is the crossover scale and $\lambda$ is the dimensionless constant. The second term in Equation (\ref{pw1}) can be considered as a power-law correction to the entropy-area law, arising from entanglement of the wave-function of the scalar field between the ground state and the exited state \cite{She}.\\
\par Motivated by these facts, we consider a non-canonical scalar field
model of the universe which is interacting with the matter sector
via some source term. The cosmological aspects of this model has
already been studied by the authors in an earlier work
\cite{sdas2}. In the present work, we investigate the validity of GSL of thermodynamics for the
non-canonical model at both the apparent horizon and the event
horizon. The validity of GSL of thermodynamics have been extensively studied for a variety of canonical scalar field models, but have not been studied for non-canonical scalar field models. As nothing is known about the nature of DE, and non-canonical scalar field models happen to be a potential candidate for DE, it is worth investigating the validity of GSL of thermodynamics for non-canonical scalar field models as well. In the present work, we consider one such toy model for interacting non-canonical scalar field \cite{sdas2} and study its thermodynamical properties. The paper is organized as follows: for the sake of
simplicity of presentation, in section II, we give a brief
overview of the cosmological model mentioned earlier \cite{sdas2}.
In section III, we check the validity of GSL of thermodynamics at
the horizon (apparent/event) assuming that first law of
thermodynamics is valid at both the horizons. In section IV, we
perform similar analysis at both apparent and event horizons
taking into account the entropy-area corrections given by
equations (\ref{log1}) and (\ref{pw1}). The last section discusses
the results.
\section{A model of non-canonical scalar field interacting with matter}
\par The most general action for a non-canonical scalar field model is given by
\be\label{action}
{\mathcal{L}} = \int\sqrt{-g} d^{4}x\left[\frac{R}{2} + {\cal L_{\phi}}(\phi,X)\right] + {\mathcal L}_{m}
\ee
where $R$ is the Ricci scalar, ${\mathcal L}_{m}$ represents the action of the background matter,
the Lagrangian density ${\cal L_{\phi}}(\phi,X)$ is an arbitrary function of the
scalar field $\phi$ such that $\phi$ is a function of time only and its kinetic
term $X$ is given by $X = \frac{1}{2}{\partial_{\mu}}\phi{\partial^{\mu}}\phi =
 \frac{1}{2}{\dot{\phi}}^2$. Here we have chosen the unit $8{\pi}G=c=1$.\\
The matter content of the universe is considered to be composed of
perfect fluid, for which the energy-momentum tensor is given by
\be T^m_{\mu\nu} = (\rho_{m} + p_{m})u_{\mu}u_{\nu} - p_{m}g_{\mu\nu}
\ee
where ${\rho}_{m}$ and $p_{m}$ are the energy density and pressure components
of the matter part of the universe respectively. The four-velocity of the
fluid is denoted by $u_{\mu}$ which satisfies $u_{\mu}u^{\mu}=1$.\\
Again, the energy-momentum tensor for the scalar field $\phi$ is
given by
\be T^{\phi}_{\mu\nu} = \frac{\partial {{\cal
L}_{\phi}}}{\partial X}{{\partial_{\mu}} \phi
{\partial_{\nu}}\phi} - g_{\mu\nu}{{\cal L}_{\phi}}. 
\ee 
From the Lagrangian density (\ref{action}), one can obtain the field
equations as \cite{sdas2} 
\be\label{e1} 
G_{\mu\nu} =
\frac{\partial {\cal{L_{\phi}}}}{\partial X}\partial_{\mu}\phi
\partial_{\nu}\phi - g_{\mu\nu}{\cal{L_{\phi}}} + T_{\mu\nu}^{m}
\ee 
and 
\be\label{e2}
{\ddot{\phi}}\left[\left(\frac{\partial
{\cal L}_{\phi}}{\partial X}\right) + (2X)\left(\frac{\partial^2
{\cal L}_{\phi}}{\partial X^2}\right)\right] + \left[3H
\left(\frac{\partial {\cal L}_{\phi}}{\partial X}\right) +
{\dot{\phi}}\left(\frac{\partial^2 {\cal L}_{\phi}}{\partial
X\partial \phi}\right)\right]{\dot{\phi}} - \left(\frac{\partial
{\cal L}_{\phi}}{\partial \phi}\right) = 0 
\ee
A number of functional forms for ${\cal{L}}_{\phi} (\phi, X)$ has been
considered in the literature. Following Fang et al \cite{Fang}, we
consider the non-canonical scalar field Lagrangian density as
\be\label{L1} {\cal L}_{\phi}(\phi,X) = F(X) - V(\phi) \ee where $F(X)$
is an arbitrary function of $X$ and $V(\phi)$ is the
self-interacting potential for the scalar field. \\
\par Following \cite{unni,  mukhanov, sdas2}, we consider $F(X) = X^2$
such that Lagrangian density for the scalar field becomes
\be\label{fX}
{\cal{L}}_{\phi}(\phi, X) = X^2 - V(\phi)~.
\ee
With (\ref{fX}), the energy density and the pressure associated with the
scalar field is obtained as \cite{sdas2}
\bea
\rho_{\phi} = 3X^2 + V(\phi)~=~\frac{3}{4}{\dot{\phi}}^4 + V(\phi),\\
p_{\phi} = X^2 - V(\phi)~=~\frac{1}{4}{\dot{\phi}}^4 - V(\phi)
\eea
We consider a homogeneous, isotropic and spatially flat FRW universe which is
characterized by the line element
\be\label{metric}
ds^{2} = dt^{2} - a^{2}(t)[dr^{2} + r^{2}d{\theta}^{2} +r^{2}sin^{2}\theta d{\phi}^{2}]
\ee
where $a{\rm{(t)}}$ is the scale factor of the universe and $t$ is the cosmic time.
Here, we only consider the spatially flat FRW universe as it is preferred by
the anisotropy of the CMBR measurement \cite{pde}.\\
So, the Einstein's field equations (\ref{e1})-(\ref{e2}) for the
space-time given by equation (\ref{metric}) takes the form
\be\label{eqx1} 
3H^{2} = {\rho}_{m} + \frac{3}{4}{\dot{\phi}}^4 +
V(\phi), 
\ee 
\be\label{eqx2} 
2{\dot{H}} + 3H^{2} =
-\frac{1}{4}{\dot{\phi}}^4 + V(\phi), 
\ee 
\be\label{eqx3}
{\dot{\rho}}_{\phi} + 3H (1 + \omega_{\phi}) \rho_{\phi} = Q, 
\ee
\be\label{eqx4} 
{\dot{\rho}}_{m} + 3H{\rho}_{m} = -Q 
\ee 
where we have considered an interaction `$Q$' between the dark matter and the scalar field component which will determine the rate of flow of energy between the two components. This type of interactions have shown to provide solution to a number of cosmological problems and resulted in a variety of dark energy models (for details one can see \cite{reint} and the references therein). Recently it has been shown that BOSS (Baryon Oscillation
Spectroscopic Survey) data also provides evidence for interacting Dark Energy \cite{boss}.  Also an interacting between the dark energy and the dark matter component provides a more general framework to work with. \\
As this work is an extension of the paper of the present authors
Das and Mamon \cite{sdas2}, following \cite{sdas2}, we consider a
simple functional form for the interaction term $Q$ as 
\be 
Q =
\zeta H {\dot{\phi}}^4, ~~~~~~~~\zeta ~~\rm{being ~a~ constant.}
\ee 
Also it is well known observationally that at present
$\omega_{\phi} \simeq -1$ \cite{vasey, davis} and for acceleration
(${\ddot{a}} > 0$) to occur, the effective equation of state (EoS)
parameter should be in the range : $-1 <\omega_{\phi} <
-\frac{1}{3}$. This indicates that the range of allowed values of
$\omega_{\phi}$ over the complete span of evolution is very small.
So, we consider that the EoS parameter $\omega_{\phi}$ as constant
(say, $\omega$) and is given by the following simple relation
 \be
\omega_{\phi} = \frac{p_{\phi}}{\rho_{\phi}} = \frac{X^2 - V}{3X^2
+ V} = \omega ~(\rm{a ~constant}) 
\ee 
With these set of simplifying assumptions, the expressions for the various
components of the non-canonical scalar field model are obtained as
: (for detailed calculations please refer to \cite{sdas2}) 
\bea
V(z) = V_{0}(1+z)^{\epsilon}, \label{V}\\
H^2 = \gamma (1+z)^{\epsilon} + B(1+z)^{3},\label{H}\\
\rho_{\phi}(z) = \frac{4}{(1 - 3\omega)}V(z),\label{rhophi}\\
p_{\phi}(z) = \frac{4\omega}{(1 - 3\omega)}V(z),\label{pphi}\\
\rho_{m}(z) = \left({3\gamma} -
\frac{4V_{0}}{1-3\omega}\right)(1+z)^{\epsilon} +
3B(1+z)^{3}\label{rhom}
 \eea
  where $\epsilon = (3 - \zeta)(1 +
\omega)$, with $V_{0}$, $B$ are positive constants\\ and $\gamma = \frac{4\omega V_{0}}{(3-\epsilon)(3\omega -1)}.$ \\
\par Using equations (\ref{rhophi}), (\ref{rhom}) and (\ref{pphi}),
one can obtain the expressions for effective energy density and
effective pressure components for the universe as
\be\label{rhoeff} 
\rho_{eff}(z) = \rho_{\phi}(z) + \rho_{m}(z) = 3
H^2 
\ee 
and 
\be\label{peff} p_{eff}(z) = p_{\phi}(z) \ee Also from
equation (\ref{H}), (\ref{eqx3}) and (\ref{eqx4}), one can obtain
the expression for $\dot{H}$ in terms of $\rho_{eff}$ and
$p_{eff}$ as \be\label{Heff} \dot{H} =
-\frac{1}{2}\left(\rho_{eff} + p_{eff}\right) 
\ee
The stability of this particular interacting non-canonical scalar field model has also been investigated in the present work \cite{sdas2} by rewriting the Einstein field equations as a plane autonomous system. It was found that for a particular choice of model parameters, there are only two physically relevant generic stable fixed points and it is classically unstable at all other points. Infact, if one computes the effective sound speed for the present model, it is found that
\be
c^{2}_{s}=\frac{\partial p_{eff}}{\partial \rho_{eff}}=\frac{\frac{\partial p_{eff}}{\partial z}}{\frac{\partial \rho_{eff}}{\partial z}}=\frac{\frac{4V_{0}\epsilon \omega}{1-3\omega}}{3\gamma\epsilon + 9B(1+z)^{3-\epsilon}}
\ee 
So, the effective sound speed crucially depends on the choice of model parameters. For the choice of parameters (consistent with the analytical model \cite{sdas2} described above), it is found that $c^{2}_{s}<0$ and the model seems to be classically unstable everywhere. However, the situation will be completely different for other choices of model parameters. With these equations, one can now check the validity of Generalized Second
Law of Thermodynamics at both apparent horizon and event horizon.
\section{Generalized Second Law of Thermodynamics on Apparent and Event Horizons}
In this section, we will check the validity of GSL of
thermodynamics on both apparent and event horizons with a prior
assumption that the first law is valid on both the horizons
\cite{nm}. From the first law of thermodynamics, one can arrive at
the expression \cite{Cai0, bousso, ujjal1} \be\label{entropy1}
T_{X}dS_{X} = -dE_{X} = 4\pi R_{X}^{3}H(\rho_{eff}+p_{eff})dt \ee
where suffix $X$ denotes either the apparent horizon $(X = A)$ or
the event horizon $(X = E)$. Similarly $T_X$ and $R_X$ denotes
the temperature and radius of the apparent/event horizons accordingly. \\
The radii of apparent and event horizons for a flat FRW universe
are defined as \cite{Cai0} 
\be\label{RA} 
R_{A}=\frac{1}{H} 
\ee 
and
\be 
R_{E}=a\int_{a}^{\infty}\frac{da}{a^{2}H} =
\frac{1}{1+z}\int_{-1}^{z}\frac{dz}{H} 
\ee 
These immediately give
\be 
\dot{R}_{A}=-\frac{\dot{H}}{H^2} 
\ee
 and
\be\label{REdot}
\dot{R}_{E}= \left(\frac{d
R_E}{dz}\right)\left(\frac{dz}{dt}\right) = \left(H R_E -
1\right). 
\ee 
Now from equation (\ref{entropy1}), one can obtain
the time rate of change of entropy function on the horizons as
\be\label{entropy2} 
T_X \dot{S}_{X}= 4 \pi R_{X}^{3} H
(\rho_{eff}+p_{eff}) 
\ee 
Again from Gibb's equation of thermodynamics \cite{Wang0} 
\be\label{intentropy}
T_{X}dS_{IX}=p_{eff}dV_X+d(E_{IX}) 
\ee
where $S_{IX}$ is the entropy inside the horizon (apparent/event), $V_X = \frac{4}{3}\pi
R_{X}^{3}$ is the volume of the horizon universe
and $E_{IX}=\rho_{eff} V_{X}$ corresponds to the internal energy. \\
From equation (\ref{intentropy}), one can obtain the time rate of
change of internal entropy function as 
\be 
T_X \dot{S}_{IX}= 4 \pi
R_{X}^{2} (\rho_{eff}+p_{eff}) \left(\dot{R}_X - H R_X\right) 
\ee
This gives the total rate of change of entropy at the horizon
(apparent / event) as 
\be\label{totentropy}
 T_{X}\dot{S}_{Xtot} =
T_{X}\dot{S}_{X} + T_{X}\dot{S}_{IX} 
\ee 
At the apparent horizon, for the present non-canonical scalar field model, equation
(\ref{totentropy}) comes out as 
\be 
T_{A}\dot{S}_{Atot} = 18 \pi
\left(\frac{\rho_{eff} + p_{eff}}{\rho_{eff}}\right)^2 
\ee
which is always positive definite irrespective of the functional forms
of
$\rho_{eff}$ and $p_{eff}$. So GSL is always valid at the apparent horizon. \\
\\
At the event horizon, it comes out as 
\be 
T_{E}\dot{S}_{Etot} =
4\pi R_{E}^{2} (\rho_{eff}+p_{eff}) \left(H R_E - 1\right) 
\ee
\begin{figure}[!h]
\centerline{\psfig{figure=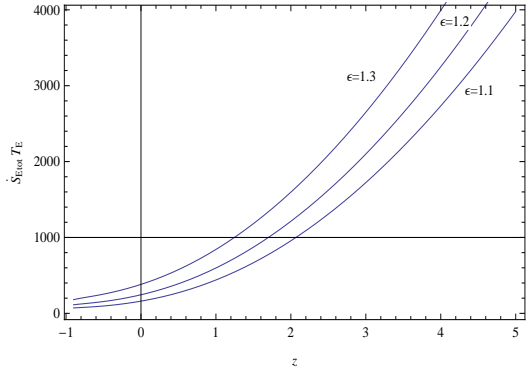,height=50mm,width=70mm}}
\caption{Plot of $T_{E}\dot{S}_{Etot}$ vs. $z$ for
different values of $\epsilon$. For this plot we have taken
$\omega = -0.9$, $V_{0} = 2$, $B=0.3$.} \label{plotFirstlawevent}
\end{figure}
\begin{figure}[!h]
\centerline{\psfig{figure=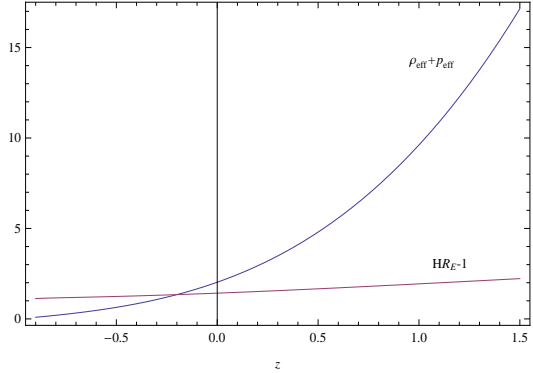,height=50mm,width=70mm}}
\caption{Plot of $\rho_{eff}+p_{eff}$ and
$HR_{E}-1$ vs  $z$ for $\epsilon=1.1$, $\omega = -0.9$, $V_{0} =
2$, $B=0.3$.} \label{plotH}
\end{figure}
At the event horizon, GSL will be valid if either (i) $ (\rho_{eff}+p_{eff}) > 0$
and $\left(H R_E - 1\right) >0$ or (ii) $ (\rho_{eff}+p_{eff}) < 0$ and
$\left(H R_E - 1\right) < 0$. Figure \ref{plotFirstlawevent} shows the plot of
$T_{E}\dot{S}_{Etot}$ vs. $z$ for different values of the model parameter
$\epsilon$. It is evident from the plot that at the event horizon GSL is valid
for each case. It has been verified that the same result is obtained for a
wide range of values of the model parameter $V_0$ as well. The reason for this
is always satisfied is for the present interacting non-canonical scalar field
toy model, always $ (\rho_{eff}+p_{eff}) > 0$ and $\left(H R_E - 1\right) >0$
which is evident from figure \ref{plotH}. Thus we can conclude that for the
present toy model, GSL is always satisfied at both event horizon and
apparent horizon.
\section{Study of Generalized Second Law of Thermodynamics on Apparent
and Event Horizons with corrected entropy}
In this section we study the validity of GSL of thermodynamics at
both the apparent horizon and event horizon for the non-canonical
scalar field model where the entropy - area relation is modified.
We have considered two types of modifications here as discussed
earlier - the logarithmic correction and the power-law correction.
The motivation behind this exercise is to check how does the GSL
of thermodynamics gets effected by these modifications for this
particular interacting non-canonical scalar field model.
\subsection{ At Apparent Horizon}
The modified entropy-area relation at the apparent horizon is
considered as 
\be\label{modentropy} 
S_A = \frac{A}{4G} = 2 \pi
f(A) ~~~~~~~~{\rm since}~8 \pi G =1~. 
\ee 
where $f(A)$ corresponds to the modification introduced whose functional form will depend
upon the type of modification (logarithmic or power-law)
incorporated and $A = 4 \pi {R_A}^2$. If $f(A) = A$, we obtain the
original entropy-area relation that has been used in the previous section. \\
Also the temperature on the apparent horizon is given by \cite{Cai0}
\be\label{modtemp}
T_A = \frac{|{\kappa}|}{2 \pi} ~= \frac{1}{2 \pi R_A}\left(1 - \frac{\dot{R}_A}{2 H R_A}\right)
= \frac{1}{4 \pi}R_A \left(\dot{H} + 2 H^2\right)
\ee
where $\kappa$ corresponds to the surface gravity term \cite{Cai0}. \\
Equations (\ref{modentropy}) and (\ref{modtemp}) along with (\ref{Heff}),
(\ref{rhoeff}) and (\ref{RA}) immediately gives
\be
T_A\dot{S}_A = 2 \pi T_A f'(A) \frac{dA}{dt} = 3 \pi f'(A) \frac{\left(\rho_{eff}
+ p_{eff}\right)}{\rho_{eff}^2}\left(\rho_{eff} - 3 p_{eff}\right)
\ee
Similarly the rate of change of entropy inside the apparent horizon is given by
\be
T_A \dot{S}_{IA} = p_{eff} \frac{d V_A}{dt} + \frac{d}{dt}\left(\rho_{eff} V_A\right)
\ee
where $V_A = \frac{4}{3}\pi {R_A}^3$ is the volume inside the apparent horizon.\\
The above equation by simple algebra takes the form
\be
T_A \dot{S}_{IA} = 6 \pi \frac{\left(\rho_{eff} + p_{eff}\right)}{\rho_{eff}^2}\left(\rho_{eff} + 3 p_{eff}\right)
\ee
This gives the total entropy at the apparent horizon as
\be\label{entropylogapp}
T_A \dot{S}_{Atot}  = 3 \pi \frac{\left(\rho_{eff} + p_{eff}\right)}{\rho_{eff}^2}\times \left[2 \left(\rho_{eff}
+ 3 p_{eff}\right) + f'(A) \left(\rho_{eff} - 3 p_{eff}\right) \right]
\ee
It is evident from equation (\ref{entropylogapp}) that for GSL to be satisfied at the
apparent horizon, the quantity $2 \left(\rho_{eff} + 3 p_{eff}\right) + f'(A) \left(\rho_{eff}
- 3 p_{eff}\right)$ has to be positive definite. However the signature of this quantity
will depend on many parameters such as $f'(A)$ and $\left(\rho_{eff} + 3 p_{eff}\right)$.
( $\left(\rho_{eff} + p_{eff}\right)$ is always positive). It has been found that for
the present model, $\left(\rho_{eff} + 3 p_{eff}\right)$ is positive at higher $z$ and
becomes negative at later stage of evolution but the value is close to $-1$ which is
not sufficiently large. So the overall scenario depends on the functional form of $f'(A)$. \\
\\
Now for logarithmic correction \cite{Meis,Jamil1,Sad1}
\bc
\bea
S_{BH}=\frac{A}{4G}+\alpha\ln \frac{A}{4G}+\beta
\frac{4G}{A}+\gamma\nonumber\\
\Rightarrow f(A) = \frac{S}{2 \pi} = A + \frac{\alpha}{2 \pi}\ln (2 \pi A) +
 \frac{\beta}{4 \pi^2 A} + \frac{\gamma}{2 \pi}\nonumber\\
\Rightarrow f'(A) = 1 + \frac{\alpha}{2 \pi}\frac{1}{A} - \frac{\beta}{4 \pi^2}\frac{1}{A^2}\label{log}
\eea
\ec
where $\alpha,~\beta$ and $\gamma$ are dimensionless constants. \\
\begin{figure}[!h]
\centerline{\psfig{figure=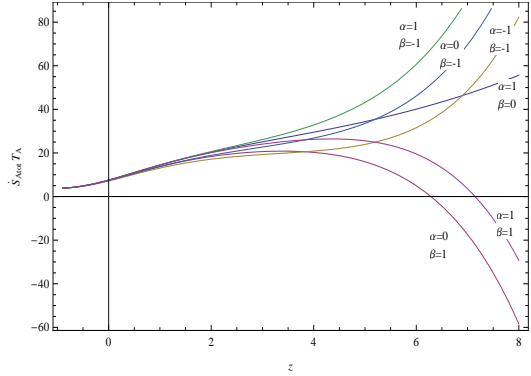,height=50mm,width=70mm}}
\caption{Plot of $T_A \dot{S}_{Atot}$ vs. $z$ for
logarithmic correction for $\epsilon=1.1$ and different
combinations of $\alpha$ and $\beta$.} \label{plotlogapparent}
\end{figure}
Figure \ref{plotlogapparent} shows that when $\beta = 0$ or
negative, GSL is always satisfied irrespective of the value of
$\alpha$. On the other hand, when $\beta$ is positive, it is found
that GSL is not satisfied at early epoch and around $z \sim 6,7$,
$T_A \dot{S}_{Atot}$ becomes positive and GSL is satisfied. The
reason for this is obvious from equation (\ref{log}). When $\beta$
is negative or $0$, $f'(A)$ remains positive (as the magnitude of
the second term in equation (\ref{log}) is very less). However for
positive $\beta$, the third term in equation (\ref{log}) dominates
at the beginning which makes the GSL invalid at the early epoch.
However for this analysis, we have kept $\alpha, \beta \sim
{\mathcal{O}}(1)$ \cite{Jamil1}.
For higher values of $\alpha, \beta$ the situation may be different. \\
\\
Similarly for power-law corrections \cite{She,Wei2}
\bea
S_{BH}=
\frac{A}{4G}\left[1-K_{\lambda}A^{1-\frac{\lambda}{2}}\right]\nonumber\\
\Rightarrow f(A) = A \left[1-K_{\lambda}A^{1-\frac{\lambda}{2}}\right]\nonumber\\
\Rightarrow f'(A) = 1 - K_{\lambda}A^{1-\frac{\lambda}{2}}\left[2
- \frac{\lambda}{2}\right]\label{pw} \eea where,
$K_{\lambda}=\frac{\lambda(4\pi)^{\frac{\lambda}{2}-1}}{(4-\lambda)r_{c}^{2-\lambda}}$,
$r_{c}$ is the crossover scale and $\lambda$ is a
dimensionless constant. \\
\begin{figure}[!h]
\centerline{\psfig{figure=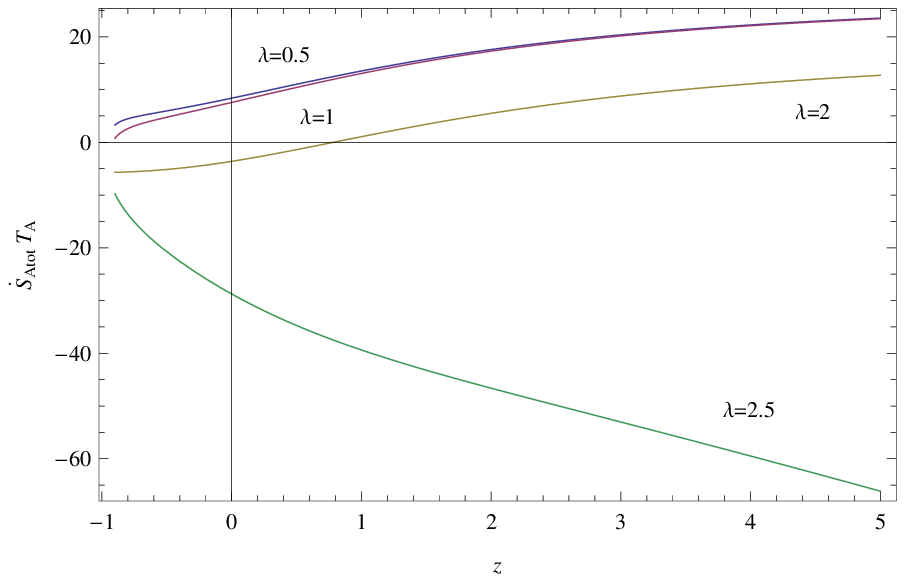,height=50mm,width=70mm}\hspace{3mm}
\psfig{figure=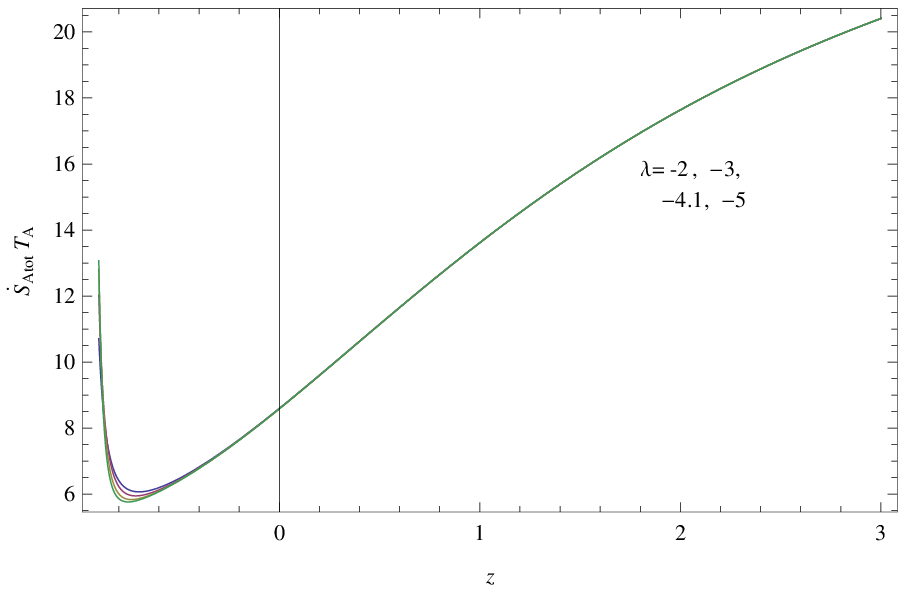,height=50mm,width=70mm}}
\caption{\normalsize{{\em Plot of $T_A \dot{S}_{Atot}$ vs. $z$ for
power-law correction for $\epsilon=1.1$ and different values of
$\lambda$. The left panel is for positive values of $\lambda$ and
the right panel is for negative values of $\lambda$.}}}
\label{plotpwapparent}
\end{figure}
Figure \ref{plotpwapparent} shows the variation of $T_A \dot{S}_{Atot}$ with $z$
for power law correction. For small positive values of $\lambda$ (say $\lambda = 0.5, 1$),
it is found that GSL is always satisfied. However as $\lambda$ increases, it is
found that GSL is satisfied at the earlier epoch and beyond $\lambda = 2.1$, GSL
is not satisfied (left panel of Figure \ref{plotpwapparent}). Also it has been found
that the plots are insensitive to the value of the cross-over scale $r_c$. On the other
hand, GSL is always satisfied for negative values of $\lambda$ (right panel of Figure \ref{plotpwapparent}). \\
So it is observed that because of the inclusion of entropy corrections, the GSL is no
longer valid throughout the evolution and crucially depends on the corrections incorporated
as well as the parameters of the toy model. This is in contrast to the result obtained in the previous section.
\subsection{ At Event Horizon}
For event horizon, we know
\bc
$T_E = \frac{1}{2 \pi R_E}~~~{\rm and}~~~R_E = \frac{1}{1+z}\int_{-1}^{z}\frac{dz}{H}$
\ec
Like before, we consider the modified entropy - area relation at the event horizon as
\bea\label{modentropy1}
S_E = 2 \pi f(A)\nonumber\\
\Rightarrow \dot{S}_E = 16 \pi^2 R_E \dot{R}_E f'(A)~.
\eea
Therefore, using (\ref{REdot})
\be
T_E\dot{S}_E = 8 \pi \dot{R}_E f'(A) = 8 \pi \left(H R_E - 1\right) f'(A)
\ee
Also, the rate of change of entropy inside the event horizon is given by
(following the same calculations as above)
\be
T_E \dot{S}_{IE} = 4 \pi R^{2}_{E} \left(\rho_{eff} + p_{eff}\right) \left[\dot{R}_E - R_E H\right]  =
- 4 \pi R^{2}_{E} \left(\rho_{eff} + p_{eff}\right)
\ee
Hence, the total entropy at the event horizon as
\be
T_E \dot{S}_{Etot} = 4 \pi R^{2}_{E} \left[\frac{2}{R^{2}_{E}}\left(H R_E - 1\right)f'(A) -
\left(\rho_{eff} + p_{eff}\right)  \right]
\ee
Now replacing $f'(A)$ from equations (\ref{log}) and (\ref{pw})
for logarithmic and power-law corrections respectively, one
arrives at the expressions for $T_E \dot{S}_{Etot}$  as function
of $z$ for both the cases.
\\
\begin{figure}[!h]
\centerline{\psfig{figure=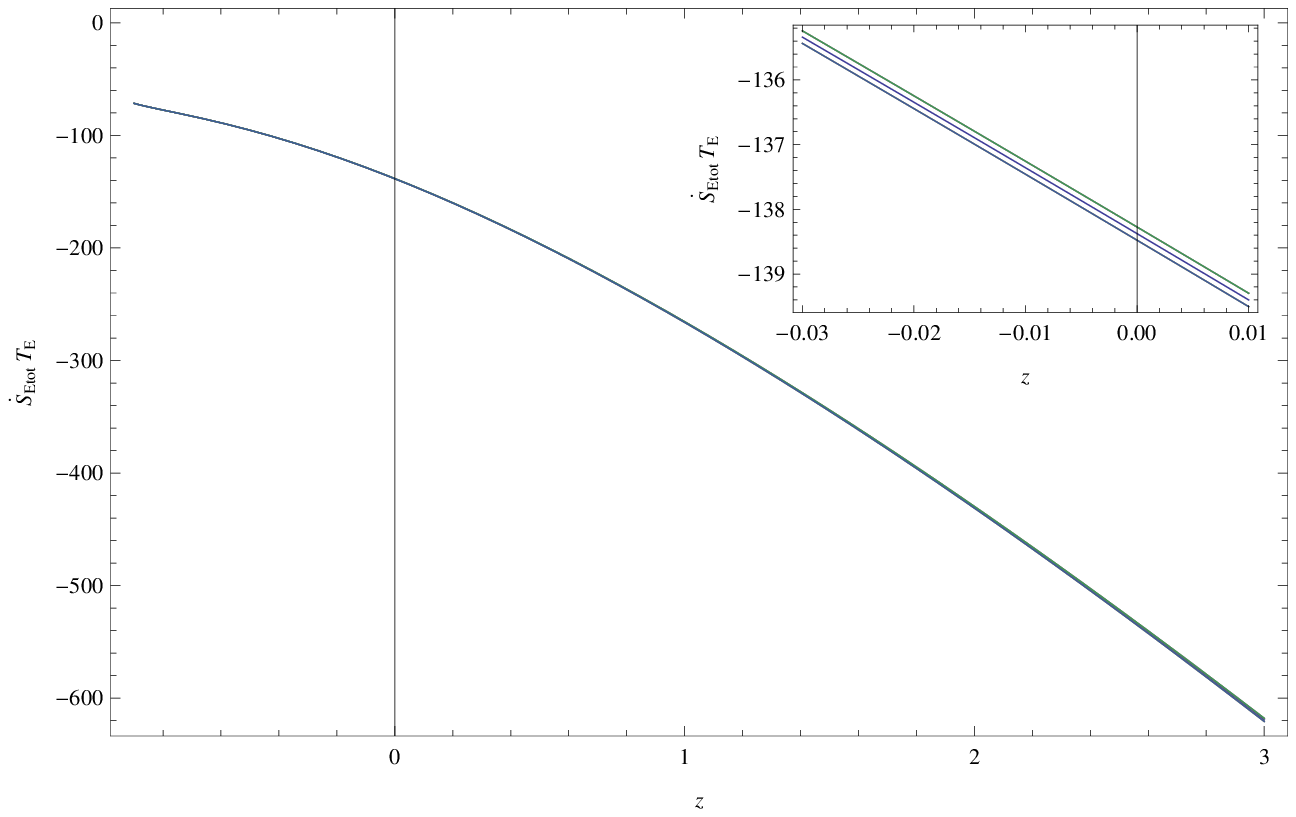,height=50mm,width=70mm}\hspace{3mm}
\psfig{figure=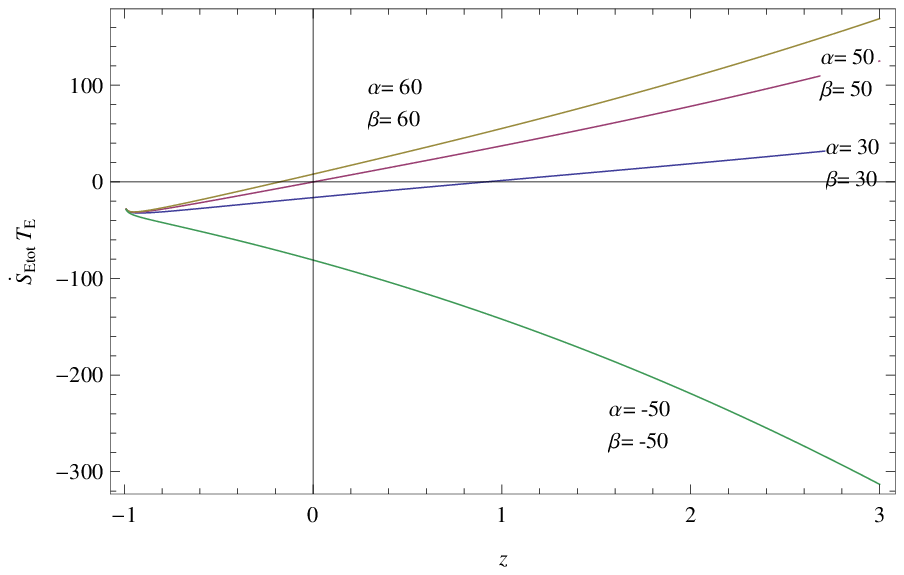,height=50mm,width=70mm}}
\caption{\normalsize{{\em Plot of $T_E \dot{S}_{Etot}$ vs. $z$ for logarithmic correction for $\epsilon=1.1$ and different values of $\alpha$ and $\beta$. The left panel is for values of $\alpha,~\beta \sim \mathcal{O}$$(1)$ and the right panel is for higher values of $\alpha,~\beta$.}}} \label{plotlogevent}
\end{figure}
Left panel of figure \ref{plotlogevent} shows that GSL is never
satisfied for this present toy model for $\alpha,~\beta \sim
\mathcal{O}$$(1)$ irrespective of the values of other model
parameters. If we consider higher positive values of $\alpha$ and
$\beta$, GSL may get satisfied however those higher values can not
be justified theoretically. So the overall conclusion that one can
derive is that for the present non-canonical scalar field model,
GSL is not satisfied at the event horizon when logarithmic
correction for entropy is taken into account.\\
\begin{figure}[!h]
\centerline{\psfig{figure=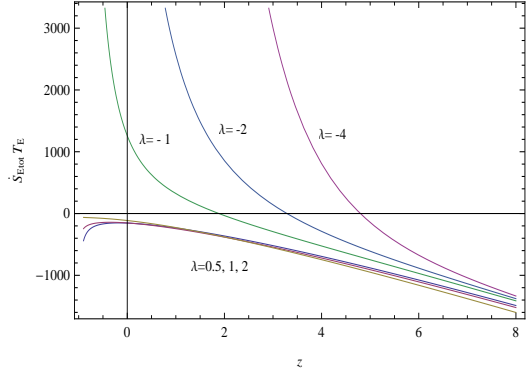,height=50mm,width=70mm}}
\caption{\normalsize{{\em Plot of $T_E \dot{S}_{Etot}$ vs. $z$ for power-law correction for $\epsilon=1.1$ and different values of $\lambda$.}}} \label{plotpwevent}
\end{figure}
Similarly figure \ref{plotpwevent} shows the plot of $T_E \dot{S}_{Etot}$ vs
$z$ for power-law  correction. It indicates that for positive values of $\lambda$,
GSL is never satisfied at the event horizon, however, for negative values of $\lambda$,
GSL is satisfied during later stages of evolution.\\
\section{Discussions}
In this work, we have considered a non-canonical scalar field
dark energy model in the framework of flat FRW background, which interacts with the dark matter sector. For the present toy model, we have assumed the phenomenological interaction term $Q = \zeta H\dot{\phi}^4$, where $\zeta$ is the interaction parameter. We have also assumed that the equation of state parameter $\omega_{\phi}=$ constant $=\omega$ and we have obtained the solutions for various cosmological quantities in terms of the redshift $z$. Using the solutions for non-canonical scalar field dark energy model, we have investigated the validity of generalized second law (GSL) of thermodynamics in various scenarios using first law and area law of thermodynamics. For this purpose, we have assumed two types of horizons viz apparent horizon and event horizon for our considered universal thermodynamics.\\
In section III, we have assumed that the first law of thermodynamics hold on apparent and event horizons. From these the rate of change of entropy have been found and using the internal entropy inside the horizons we have found the rate of change of total entropies of the Universe bounded by the apparent and event horizons. If the Universe is bounded by apparent horizon, we can see that the GSL of thermodynamics is always satisfied irrespective of the other parameters. On the other hand, for event horizon, the rate of change of total entropy has been calculated in term of $\rho_{eff},~p_{eff},~H$ and $R_{E}$. We have observed that at the event horizon, GSL is valid as both $
(\rho_{eff}+p_{eff}) > 0$ and $\left(H R_E - 1\right) >0$. The results are found to be valid for a wide range of values of the model parameter as well. Thus we can conclude that for the present toy model, GSL is always satisfied at both event horizon and apparent horizon.\\
In section IV, we have assumed the modified entropy-area relation at the apparent horizon and event horizon. On the apparent horizon, we have taken the temperature as\\ $\frac{1}{2\pi R_{A}} {\left(1- \frac{{\dot{R}}_{A}}{2HR_{A}}\right)}$. It has been found that for GSL to be satisfied at the apparent horizon, the quantity $2 \left(\rho_{eff} + 3 p_{eff}\right) + f'(A) \left(\rho_{eff} - 3 p_{eff}\right)$ has to be positive definite. However the signature of this quantity will depend on many parameters such as $f'(A)$ and $\left(\rho_{eff} + 3p_{eff}\right)$. It has been found that for the present model,
$\left(\rho_{eff} + 3p_{eff}\right)$ is positive at higher $z$ and becomes negative at later stage of evolution but the value is close to $-1$ which is not sufficiently large. So the overall scenario depends on the functional form of $f'(A)$.
\par On the apparent horizon, we have assumed two types of corrected
entropies, viz, logarithmic and power law entropies. We have found that for logarithmic correction, the validity of GSL crucially depends on the values of the correction parameters $\alpha$ and $\beta$. When $\beta=0$ or negative, GSL is always satisfied irrespective of value of $\alpha$, but for $\beta > 0$ it has been found that GSL is satisfied at late time. When $\beta$ is negative or $0$, $f'(A)$ remains positive. However for positive $\beta$, the third term in equation (\ref{log}) dominates at the beginning which makes the GSL invalid at the early epoch. However for this analysis, we have kept $\alpha, \beta \sim {\mathcal{O}}(1)$. For higher values of $\alpha, \beta$ the situation may be different. For power-law corrections also, it has been found that the validity of the GSL depends on the correction parameter $\lambda$ such that for small positive values of $\lambda$, GSL is satisfied always but with increase in the value of $\lambda$, GSL is not satisfied. However, GSL is always satisfied for negative values of $\lambda$. So it is observed that because of the inclusion of entropy corrections, the GSL is no longer valid throughout the evolution and crucially depends on the corrections incorporated as well as the parameters of the toy model.
\par We have also applied the logarithmic and power law corrections of entropies on event
horizon as well. It has been found that for logarithmic correction GSL is never satisfied at the event horizon for this present toy model for $\alpha,~\beta \sim \mathcal{O}$$(1)$ irrespective of the values of other model parameters. If we consider higher positive
values of $\alpha$ and $\beta$, GSL may get satisfied however those higher values can not be justified theoretically. So the overall conclusion that one can derive is that for the present non-canonical scalar field model, GSL is not satisfied at the event horizon when logarithmic correction for entropy is taken into account. However, for power-law correction it is seen that for positive values of $\lambda$, GSL is never satisfied at the event horizon, however, for negative values of $\lambda$, GSL is satisfied during later stages of
evolution.\\ \\
{\bf Acknowledgement:}\\ \\
The authors (SD and UD) are thankful to IUCAA, Pune, India for
warm hospitality where part of the work was carried out. AAM acknowledges UGC, Govt.
of India for financial support through Maulana Azad National Fellowship.\\


\begin{thebibliography}{25}
\bibitem{Riess} A. G. Riess et al. (Supernova Search Team
Collaboration), Astron. J. {\bf116}, 1009 (1998).
\bibitem{Perl} S. Perlmutter et al. (Supernova Cosmology Project Collaboration),
Astrophys. J. {\bf517}, 565 (1999).
\bibitem{Teg} M. Tegmark et al. (SDSS Collaboration), Phys. Rev. D {\bf69},
103501 (2004).
\bibitem{Abaz} K. Abazajian et al. (SDSS
Collaboration), Astron. J. {\bf128}, 502 (2004); Astron. J. 129, 1755
(2005).
\bibitem{Sper1} D. N. Spergel et al. (WMAP Collaboration), Astrophys. J.
Suppl. Ser. {\bf148}, 175 (2003).
\bibitem{Sper2} D. N. Spergel et al., Astrophys. J. Suppl. {\bf170}, 377 (2007).
\bibitem{quin}V. Sahni, astro-ph/0403324.
\bibitem{martin}J. Martin, Mod. Phys. Lett. A {\bf 23}, 1252 (2008).
\bibitem{Tsuji}S. Tsujikawa, Class. Quantum Grav. {\bf 30}, 21400 (2013).
\bibitem{sdas2}S. Das and A. A. Mamon, Astrophys. Space Sci., {\bf 355}, 371 (2015) [arXiv:1407.1666 [gr-qc]].
\bibitem{Fang}W. Fang, H. Q. Lu and Z. G. Huang, Class. Quant. Grav. {\bf 24}, 3799 (2007).
\bibitem{unni} S. Unnikrishnan et al., arXiv:1205.0786 [astro-ph.CO].
\bibitem{Berg} M. S. Berger, H. Shojaei, Phys. Rev. D {\bf 74}, 043530 (2006).

\bibitem{Ca} R.-G. Cai, A. Wang, JCAP, {\bf 03}, 002 (2005).
\bibitem{Zim} W. Zimdahl, Int. J. Mod. Phys. D 142319 (2005).
\bibitem{Hu} B. Hu, Y. Ling, Phys. Rev. D {\bf 73} 123510 (2006).
\bibitem{msami}E. J. Copeland, M. Sami, S. Tsujikawa, Int.J.Mod.Phys. D {\bf 15}, 1753 (2006).
\bibitem{boss}E. Abdalla et al., arXiv:1412.2777 [astro-ph.CO].
\bibitem{Haw} S. W. Hawking, Commun. Math. Phys. {\bf 43}, 199 (1975).
\bibitem{Bek} J. D. Bekenstein, Phys. Rev. D {\bf 7}, 2333 (1973).
\bibitem{Bar} J. M. Bardeen, B. Carter and S. W. Hawking, Commun. Math. Phys., {\bf 31}, 161 (1973).
\bibitem{Jacob} T. Jacobson, Phys. Rev. Lett., {\bf 75}, 1260 (1995).
\bibitem{Pad} T. Padmanabhan, Class. Quantum Grav., {\bf 19}, 5387 (2002).
\bibitem{Cai0} R. -G. Cai, S. P. Kim, JHEP, {\bf 0502}, 050 (2005).
\bibitem{Gibb} G. W. Gibbons and S. W. Hawking, Phys. Rev. D {\bf 15}, 2738 (1977).
\bibitem{Wang0} B. Wang, Y. G. Gong and E. Abdalla, Phys. Rev. D {\bf 74}, 083520 (2006).
\bibitem{Set0} M. R. Setare, and S. Shafei, JCAP, {\bf 09}, 011 (2006).
\bibitem{Iz} G. Izquierdo and D. Pavon, Phys. Rev. D {\bf 70}, 127505 (2004).
\bibitem{Frol} A. V. Frolov and L. Kofman, JCAP, {\bf 0305}, 009 (2003).
\bibitem{Gong} Y. Gong, B. Wang and A. Wang, JCAP, {\bf 01}, 024 (2007).
\bibitem{Shey} A. Sheykhi and B. Wang, Phys. Lett. B {\bf 678}, 434 (2009).
\bibitem{Wang1} A. Wang and Y. Wu, JCAP, {\bf 0907}, 012 (2009).
\bibitem{Ak} M. Akbar and R. -G. Cai, Phys. Lett. B {\bf 635}, 7 (2006).
\bibitem{Sad} H. M. Sadjadi, Phys. Rev. D {\bf 76}, 104024 (2007).
\bibitem{Zhu} T. Zhu and J-R. Ren, Eur. Phys. J. C {\bf 62}, 413 (2009).
\bibitem{Cai00} R-G. Cai et al, Class.Quant.Grav., {\bf 26}, 155018 (2009).
\bibitem{Meis} K. A. Miessner, {\it Class. Quantum Grav.} {\bf 21}
5245 (2004).
\bibitem{Jamil1} M. Jamil and M. U. Farooq, {\it JCAP} {\bf 03}
001 (2010).
\bibitem{Sad1} H. M. Sadjadi and M. Jamil, {\it Europhys. Lett.}
{\bf 92}, 69001 (2010).
\bibitem{sd3c}S. Das, S. Shankaranarayanan, S. Sur, Phys. Rev. D {\bf 77}, 064013 (2008).
\bibitem{She} A. Sheykhi and M. Jamil, Gen. Rel.
Grav., {\bf 43}, 2661 (2011).
\bibitem{Wei2} H. Wei, Commun. Theor. Phys. 52, 743 (2009).
\bibitem{mukhanov}V. Mukhanov et al., JCAP {\bf 0602}, 004 (2006).
\bibitem{pde}P. de Bernardis et al., Nature, {\bf 400}, 955, (2000).
\bibitem{reint}Y. L. Bolotin et al., arXiv:1310.0085 [astro-ph.CO].
\bibitem{vasey}W. M. Wood-Vasey et al.: Astrophys. J. {\bf 666}, 694 (2007).
\bibitem{davis}T. M. Davis et al., Astrophys. J. {\bf 666}, 716 (2007).
\bibitem{nm}N. Mazumder and S. Chakraborty, Class. Quant. Grav. {\bf 26}, 195016 (2009).
\bibitem{bousso}R. S. Bousso, Phys. Rev. D {\bf 71}, 064024 (2005).
\bibitem{ujjal1}S. Bhattacharya and U. Debnath, Can. J. Phys. {\bf 89}, 883 (2011).
\end{thebibliography}
\end{document}